\documentclass{article}
\usepackage{amsmath,amsfonts,graphicx,mlspconf}
\usepackage{booktabs}
\usepackage{flushend}
\usepackage{cite}
\usepackage{url}

\copyrightnotice{Preprint submitted to IEEE MLSP 2023}

\title{Improved Vocal Effort Transfer Vector Estimation for Vocal Effort-Robust Speaker Verification}
\name{Iv\'an L\'opez-Espejo$^{1,2,*}$, Santi Prieto$^{3,*}$, Alfonso Ortega$^4$, Eduardo Lleida$^4$\thanks{$^*$These authors contributed equally to this work.\\This work has been funded, in part, by the European Union's Horizon 2021 research and innovation program under the Marie Skłodowska-Curie grant agreement No. 101062614. Also in part, this work has been supported by the European Union's Horizon 2020 research and innovation program under the Marie Skłodowska-Curie grant agreement No. 101007666, MCIN/AEI/10.13039/501100011033, and the European Union NextGenerationEU/PRTR under Grants PDC2021-120846-C41 and PID2021-126061OB-C44.}}
\address{$^1$Department of Electronic Systems, Aalborg University, Denmark\\
          $^2$Center for Robust Speech Systems (CRSS), The University of Texas at Dallas, USA\\
          $^3$VeriDas $\vert$ das-Nano, Navarre, Spain\\
		 $^4$ViVoLab, Arag\'on Institute for Engineering Research (I3A)\\
          University of Zaragoza, Spain\\
		 \texttt{\small ivl@es.aau.dk, sprieto@veridas.com, \{ortega,lleida\}@unizar.es}
}

\begin{document}
%\ninept

\maketitle

\begin{abstract}
Despite the maturity of modern speaker verification technology, its performance still significantly degrades when facing non-neutrally-phonated (e.g., shouted and whispered) speech. To address this issue, in this paper, we propose a new speaker embedding compensation method based on a minimum mean square error (MMSE) estimator. This method models the joint distribution of the vocal effort transfer vector and non-neutrally-phonated embedding spaces and operates in a principal component analysis domain to cope with non-neutrally-phonated speech data scarcity. Experiments are carried out using a cutting-edge speaker verification system integrating a powerful self-supervised pre-trained model for speech representation. In comparison with a state-of-the-art embedding compensation method, the proposed MMSE estimator yields superior and competitive equal error rate results when tackling shouted and whispered speech, respectively.
\end{abstract}
\begin{keywords}
Speaker verification, vocal effort, embedding compensation, shouted speech, whispered speech
\end{keywords}
\section{Introduction}
\label{sec:intro}

State-of-the-art speaker verification technology achieves impressive performance when dealing with neutrally-phonated (i.e., normal) speech \cite{ECAPATDNN20,Chen22,WavLM}. However, because normal speech data are mostly used ---due to obvious reasons--- to train speaker verification systems, their performance tends to dramatically drop in the presence of non-neutrally-phonated (e.g., shouted and whispered) speech \cite{Prieto20,Prieto22}. To mitigate this issue, previous work \cite{Prieto20,Prieto22} explored a series of minimum mean square error (MMSE) techniques estimating normal speaker embeddings from non-neutrally-phonated ones. Among all of these techniques, multi-environment model-based linear normalization (\emph{MEMLIN}) \cite{MEMLIN} ---modeling both the normal and non-neutrally-phonated embedding spaces by Gaussian mixtures--- provided the best performance in terms of equal error rate (EER) when dealing with both shouted and whispered speech \cite{Prieto22}.

Under Gaussian mixture modeling assumption, it is well known that the MMSE estimator can be expressed as a weighted sum of a set of partial estimates \cite{JAGL13}. A shortcoming of MEMLIN is that the set of partial estimates is pre-computed (during an offline training stage) and fixed \cite{MEMLIN}. Therefore, these partial estimates do not account for the specificities of the non-neutrally-phonated embeddings observed at test time. In this paper, we propose an alternative MMSE estimator that overcomes this MEMLIN's limitation by modeling the joint distribution of the vocal effort transfer vector\footnote{As explained in Section \ref{sec:mmse}, the vocal effort transfer vector relates equivalent normal and non-neutrally-phonated speaker embeddings according to an additive model.} and non-neutrally-phonated embedding spaces. Furthermore, to circumvent non-neutrally-phonated speech data scarcity, we also propose to carry out the estimation in a principal component analysis (PCA) domain. In fact, this \emph{data scarcity prevents us from leveraging deep learning for embedding compensation}.

We conduct experiments employing a state-of-the-art speaker verification system consisting of the concatenation of a powerful self-supervised pre-trained model for speech representation so-called WavLM \cite{WavLM} and an ECAPA-TDNN \cite{ECAPATDNN20} back-end for speaker embedding extraction. In comparison with MEMLIN, the proposal at hand shows superior and competitive EER performance when dealing with shouted and whispered speech, respectively.

The remainder of this manuscript is organized as follows. Our normal speaker embedding estimation methodology is developed in Section \ref{sec:mmse}. Section \ref{sec:overview} provides the reader with an overview of the whole vocal effort-robust speaker verification system. Section \ref{sec:results} is devoted to discuss the experimental results. Finally, Section \ref{sec:conclusions} wraps up this work.

\section{Normal Speaker Embedding Estimation}
\label{sec:mmse}

\subsection{Problem Statement}
\label{ssec:statement}

Given a particular speaker, let $\tilde{\mathbf{x}}\in\mathbb{R}^D$ be a $D$-dimensional speaker embedding extracted from an utterance with normal vocal effort. Furthermore, let $\tilde{\mathbf{y}}\in\mathbb{R}^D$ be a non-neutrally-phonated counterpart of $\tilde{\mathbf{x}}$. Then, we assume the following additive model:
\begin{equation}
    \tilde{\mathbf{y}}=\tilde{\mathbf{x}}+\tilde{\mathbf{v}},
    \label{eq:additive}
\end{equation}
where $\tilde{\mathbf{v}}\in\mathbb{R}^D$ represents a \emph{vocal effort transfer vector} between the normal and non-neutrally-phonated modes.

For vocal effort-robust speaker verification purposes, previous work \cite{Prieto20,Prieto22} proposed to estimate $\tilde{\mathbf{x}}$ from an estimate of the vocal effort transfer vector, $\hat{\tilde{\mathbf{v}}}$, by following the additive model of Eq. (\ref{eq:additive}):
\begin{equation}
    \hat{\tilde{\mathbf{x}}}=\tilde{\mathbf{y}}-\hat{\tilde{\mathbf{v}}}.
    \label{eq:x_estimation}
\end{equation}
Several MMSE compensation techniques where studied in \cite{Prieto20,Prieto22} to realize Eq. (\ref{eq:x_estimation}), and, among them, MEMLIN \cite{MEMLIN} showed to be the best performing one. Assuming that the non-neutrally-phonated embedding domain is modeled by a $K$-component Gaussian mixture model (GMM), MEMLIN ---which also models the normal embedding space by another $K$-component GMM--- approximates $\tilde{\mathbf{x}}$ from a weighted combination of $K$ \emph{partial estimates} $\left\{\hat{\tilde{\mathbf{v}}}^{\{k\}};\;k=1,...,K\right\}$:
\begin{equation}
    \hat{\tilde{\mathbf{x}}}=\tilde{\mathbf{y}}-\underbrace{\displaystyle\sum_{k=1}^K P(k|\tilde{\mathbf{y}})\hat{\tilde{\mathbf{v}}}^{\{k\}}}_{\mbox{$\hat{\tilde{\mathbf{v}}}$}},
    \label{eq:mmsegmm_base}
\end{equation}
where $\left\{P(k|\tilde{\mathbf{y}});\;k=1,...,K\right\}$ are the combination weights.

As introduced in Section \ref{sec:intro}, a limitation of MEMLIN is that, unlike the combination weights, the set of partial estimates $\left\{\hat{\tilde{\mathbf{v}}}^{\{k\}};\;k=1,...,K\right\}$ is independent of the observed non-neutrally-phonated embedding $\tilde{\mathbf{y}}$, which constrains the potentials of the Bayesian estimation framework\footnote{In MEMLIN, the set of partial estimates is pre-computed during an offline training stage \cite{MEMLIN}.}. In the next subsection, we propose a different MMSE compensation approach where also the partial estimates exploit $\tilde{\mathbf{y}}$, which yields significant speaker verification performance improvements in Section \ref{sec:results}.

\subsection{Estimation Methodology}
\label{ssec:method}

Inspired by classical noise-robust speech recognition methods such as front-end joint uncertainty decoding (FE-Joint) \cite{Liao08} and stereo-based stochastic mapping (SSM) \cite{Afify09}, we explore jointly modeling the vocal effort transfer vector and non-neutrally-phonated embedding domains by means of a $K$-component GMM $p(\tilde{\mathbf{z}}=(\tilde{\mathbf{v}},\tilde{\mathbf{y}}))$. However, estimating $p(\tilde{\mathbf{z}}=(\tilde{\mathbf{v}},\tilde{\mathbf{y}}))$ requires the computation of $2D\times 2D$ covariance matrices that are \emph{ill-conditioned} under our non-neutrally-phonated speech data scarcity scenario (see Subsection \ref{ssec:corpora}). To deal with this issue, we propose to use PCA as detailed below.

Let $\mathbf{W}_L$ be a $D\times L$ PCA transform matrix calculated from normal and non-neutrally-phonated embeddings. This matrix is comprised, column-wise, of $L$ principal eigenvectors, where $L\ll D$. We can express $\tilde{\mathbf{v}}$ and $\tilde{\mathbf{y}}$ in the PCA domain as
\begin{equation}
\mathbf{v}=\mathbf{W}_L^\top\tilde{\mathbf{v}},\hspace{1cm}\mathbf{y}=\mathbf{W}_L^\top\tilde{\mathbf{y}}.
\end{equation}
Then, the joint variable $\mathbf{z}=(\mathbf{v},\mathbf{y})$, $\mathbf{z}\in\mathbb{R}^{2L}$, is modeled by a $K$-component GMM:
\begin{equation}
p\left(\mathbf{z}\right)=\displaystyle\sum_{k=1}^{K}P(k)\mathcal{N}\left(\mathbf{z}\left|\boldsymbol\mu_z^{\{k\}},\boldsymbol\Sigma_z^{\{k\}}\right)\right..
\label{eq:gmm}
\end{equation}
In Eq. (\ref{eq:gmm}), $\left\{P(k);\;k=1,...,K\right\}$ is the set of prior probabilities, whereas the mean vector $\boldsymbol\mu_z^{\{k\}}$ and covariance matrix $\boldsymbol\Sigma_z^{\{k\}}$ of the $k$-th Gaussian density can be partitioned as
\begin{equation}
    \boldsymbol\mu_z^{\{k\}} = \left(\begin{array}{c}
      \boldsymbol\mu_v^{\{k\}} \\
      \boldsymbol\mu_y^{\{k\}}
    \end{array}\right),\;\;\;
    \boldsymbol\Sigma_z^{\{k\}} = \left(\begin{array}{cc}
      \boldsymbol\Sigma_{vv}^{\{k\}} & \boldsymbol\Sigma_{vy}^{\{k\}} \\
      \boldsymbol\Sigma_{yv}^{\{k\}} & \boldsymbol\Sigma_{yy}^{\{k\}}
    \end{array}\right),
    \label{eq:params}
\end{equation}
where all $\boldsymbol\Sigma_{vv}^{\{k\}}$, $\boldsymbol\Sigma_{vy}^{\{k\}}=\left(\boldsymbol\Sigma_{yv}^{\{k\}}\right)^\top$ and $\boldsymbol\Sigma_{yy}^{\{k\}}$ are $L\times L$ diagonal matrices.

Given Eq. (\ref{eq:gmm}), the MMSE estimate of $\mathbf{v}$, $\hat{\mathbf{v}}$, is calculated as follows:
\begin{equation}
\begin{array}{lll}
      \hat{\mathbf{v}} & = & \mathbb{E}(\mathbf{v}|\mathbf{y}) = \displaystyle\int_{\mathbf{v}} \mathbf{v}p(\mathbf{v}|\mathbf{y})d\mathbf{v}\vspace{0.15cm}\\
      & = & \displaystyle\sum_{k=1}^K\int_{\mathbf{v}} \mathbf{v}p\left(\mathbf{v},k|\mathbf{y}\right)d\mathbf{v}\vspace{0.15cm}\\
      & = & \displaystyle\sum_{k=1}^K P(k|\mathbf{y})\int_{\mathbf{v}}\mathbf{v}p(\mathbf{v}|\mathbf{y},k)d\mathbf{v}\vspace{0.15cm}\\
      & = & \displaystyle\sum_{k=1}^KP(k|\mathbf{y})\underbrace{\mathbb{E}\left(\mathbf{v}\left|\mathbf{y},k\right)\right.}_{\mbox{$\hat{\mathbf{v}}^{\{k\}}$}},
\end{array}
\label{eq:mmse_gmm}
\end{equation}
where $\mathbb{E}(\cdot)$ denotes the expectation operator. On the one hand, the combination weights in Eq. (\ref{eq:mmse_gmm}) are obtained, by means of the Bayes' rule, according to
\begin{equation}
    P(k|\mathbf{y})=\frac{p(\mathbf{y}|k)P(k)}{\sum_{k'=1}^Kp(\mathbf{y}|k')P(k')},\;\;\;\;\;k=1,...,K,
    \label{eq:comb_weights}
\end{equation}
where $p(\mathbf{y}|k)=\mathcal{N}\left(\mathbf{y}\left|\boldsymbol\mu_y^{\{k\}},\boldsymbol\Sigma_{yy}^{\{k\}}\right)\right.$. On the other hand, given that the joint density $p(\mathbf{z}=(\mathbf{v},\mathbf{y})|k)$ is Gaussian, the conditional density $p(\mathbf{v}|\mathbf{y},k)$ is also Gaussian, and, therefore, $\mathbb{E}\left(\mathbf{v}\left|\mathbf{y},k\right)\right.$, i.e., the partial estimates in Eq. (\ref{eq:mmse_gmm}), can be expressed, $\forall k\in\{1,...,K\}$, as \cite{Papoulis02}
\begin{equation}
    \mathbb{E}\left(\mathbf{v}\left|\mathbf{y},k\right)\right. = \boldsymbol\mu_v^{\{k\}}+\boldsymbol\Sigma_{vy}^{\{k\}}\left(\boldsymbol\Sigma_{yy}^{\{k\}}\right)^{-1}\left(\mathbf{y}-\boldsymbol\mu_y^{\{k\}}\right).
\label{eq:partial}
\end{equation}

Finally, an estimate of the normal embedding $\tilde{\mathbf{x}}$ is achieved by means of Eq. (\ref{eq:x_estimation}) along with the application of the inverse PCA transform to the result of Eq. (\ref{eq:mmse_gmm}), namely,
\begin{equation}
    \hat{\tilde{\mathbf{x}}}=\tilde{\mathbf{y}}-\underbrace{\mathbf{W}_L\hat{\mathbf{v}}}_{\mbox{$\hat{\tilde{\mathbf{v}}}$}}.
\end{equation}

Note that, in order to apply this method in Section \ref{sec:results}, both the PCA transform matrix $\mathbf{W}_L$ and the GMM $p(\mathbf{z})$ are calculated from a training set comprising paired normal and non-neutrally-phonated embeddings (see Subsection \ref{ssec:corpora}).

For the sake of reproducibility, a Python implementation of this speaker embedding compensation methodology has been made publicly available\footnote{\url{https://ilopezes.files.wordpress.com/2023/06/mmsev.zip}}.

\section{System Overview}
\label{sec:overview}

\begin{figure}
    \centering
    \begin{picture}(100,280)
    \put(-75,-10){\includegraphics[width=\linewidth]{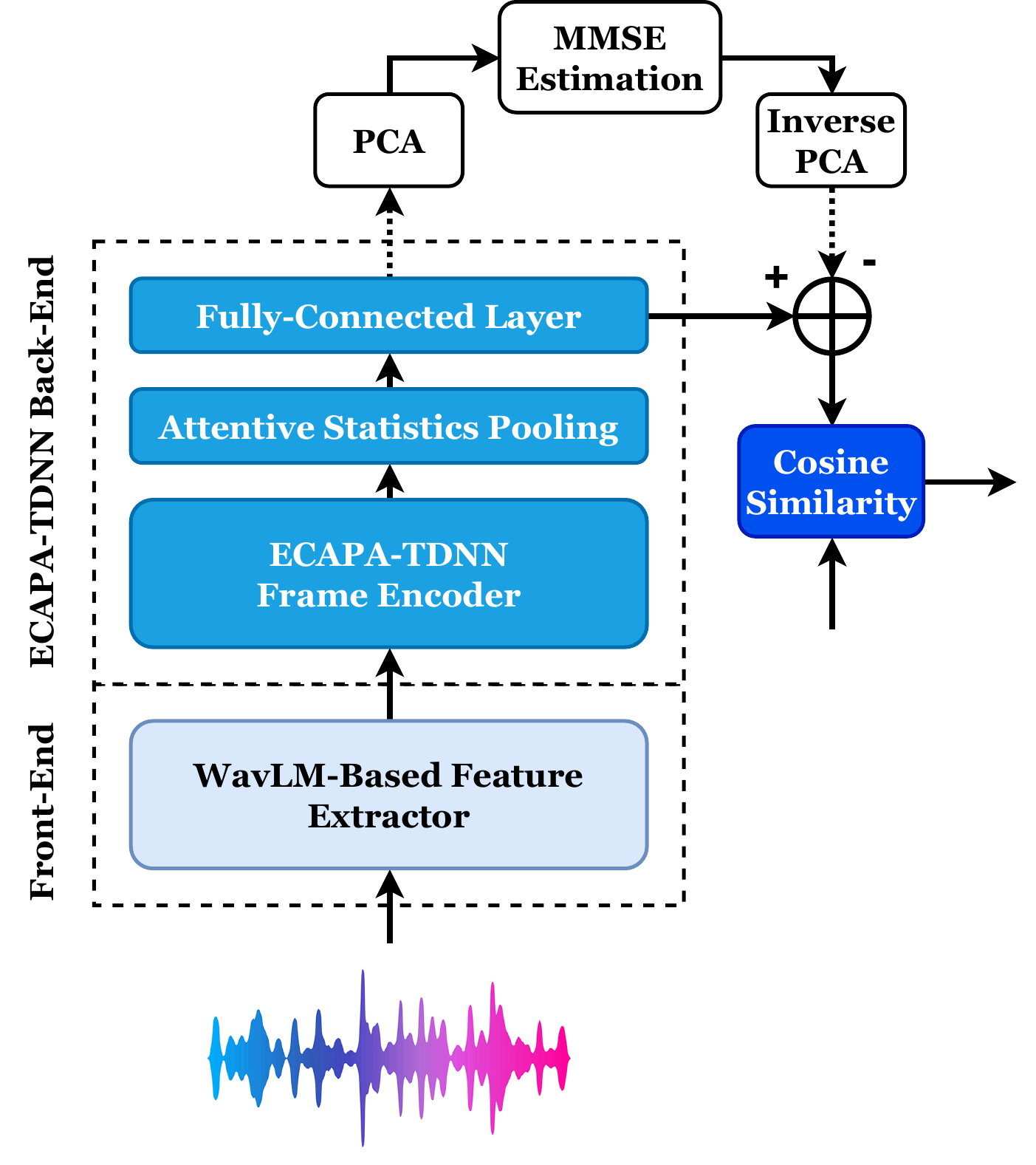}}
    \put(5,212){$\tilde{\mathbf{y}}$}
    \put(5,247){$\mathbf{y}$}
    \put(97,254){$\hat{\mathbf{v}}$}
    \put(124,210){$\hat{\tilde{\mathbf{v}}}$}
    \put(97,196){$\tilde{\mathbf{y}}$}
    \put(124,172){$\hat{\tilde{\mathbf{x}}}$}
    \put(114,106){$\tilde{\mathbf{x}}_{\mbox{\scriptsize ref}}$}
    \put(146,156){$s_c$}
    \end{picture}
    \caption{Block diagram of the proposed vocal effort-robust speaker verification system. See the text for further details.}
    \label{fig:system}
\end{figure}
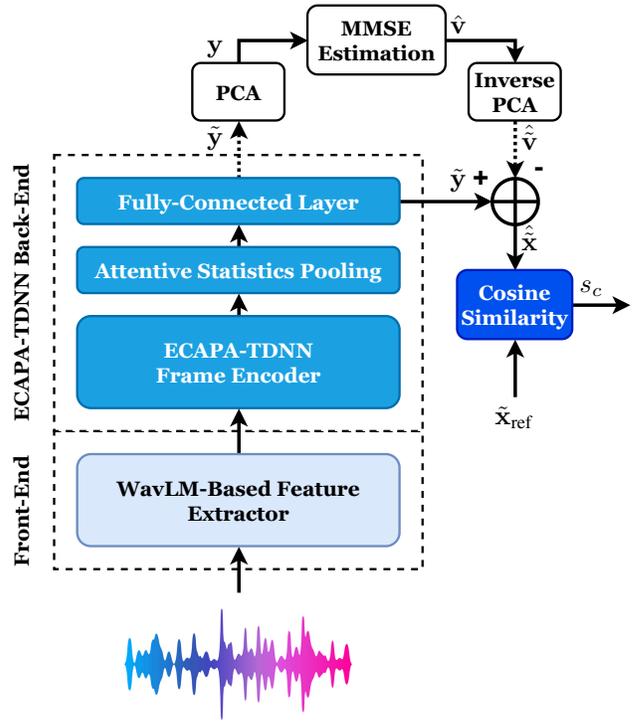

Figure \ref{fig:system} depicts a block diagram of the proposed vocal effort-robust speaker verification system. First, the powerful self-supervised pre-trained model WavLM \cite{WavLM} is used to compute a high-level representation of the input speech signal. Based on a Transformer structure, WavLM extends HuBERT \cite{HuBERT} to masked speech prediction and de-noising to allow the pre-trained model to perform well in a variety of speech processing tasks including speaker verification. Second, an ECAPA-TDNN \cite{ECAPATDNN20} back-end extracts a speaker embedding from the representation outputted by WavLM. Then, the speaker embedding compensation methodology of Section \ref{sec:mmse} is applied only in the case that the embedding comes from non-neutrally-phonated speech. To detect this case, a simple, yet virtually flawless logistic regression-based detector \cite{Prieto20,Prieto22} can be used. That being said, note that the results reported in Section \ref{sec:results} are obtained by \emph{oracle} non-neutrally-phonated speech detection for the sake of simplicity. Finally, the resulting embedding is compared with a reference embedding $\tilde{\mathbf{x}}_{\mbox{\scriptsize ref}}$ by cosine similarity to produce a score $s_c$.

\subsection{Shouted and Whispered Speech Corpora}
\label{ssec:corpora}

For experimental purposes, we consider the vocal effort modes shouted and whispered in addition to normal. To this end, we employ two different (i.e., disjoint) corpora: the speech corpus informed in \cite{Cemal13}, which comprises paired shouted-normal speech utterances in Finnish from 22 speakers, and CHAINS (CHAracterizing INdividual Speakers) \cite{Fred06}, which contains paired whispered-normal speech utterances in English from 36 speakers. Due to speech data scarcity, all the embedding compensation experiments in Section \ref{sec:results} are performed ---as in \cite{Prieto22}--- by following a leave-one-speaker-out cross-validation strategy, which serves to split the corpora into training and test sets.

We consider the following 4 test conditions (trial lists) under the shouted-normal scenario: \textbf{A$_{\mbox{s}}$-A$_{\mbox{s}}$} (all shouted and normal utterances \emph{vs}. all shouted and normal utterances; 557,040 trials), \textbf{N$_{\mbox{s}}$-N$_{\mbox{s}}$} (normal utterances \emph{vs}. normal utterances; 139,128 trials), \textbf{S-S} (shouted utterances \emph{vs}. shouted utterances; 139,128 trials) and \textbf{N$_{\mbox{s}}$-S} (normal utterances \emph{vs}. shouted utterances; 278,784 trials). Furthermore, we similarly examine 4 equivalent test conditions under the whispered-normal scenario, namely, \textbf{A$_{\mbox{w}}$-A$_{\mbox{w}}$} (2,821,498 trials), \textbf{N$_{\mbox{w}}$-N$_{\mbox{w}}$} (705,078 trials), \textbf{W-W} (704,950 trials) and \textbf{N$_{\mbox{w}}$-W} (1,411,344 trials).

For further details about these corpora, the reader is referred to \cite{Cemal13,Fred06} and \cite{Prieto22}.

\begin{figure*}
    \centering
    \includegraphics[width=0.9\linewidth]{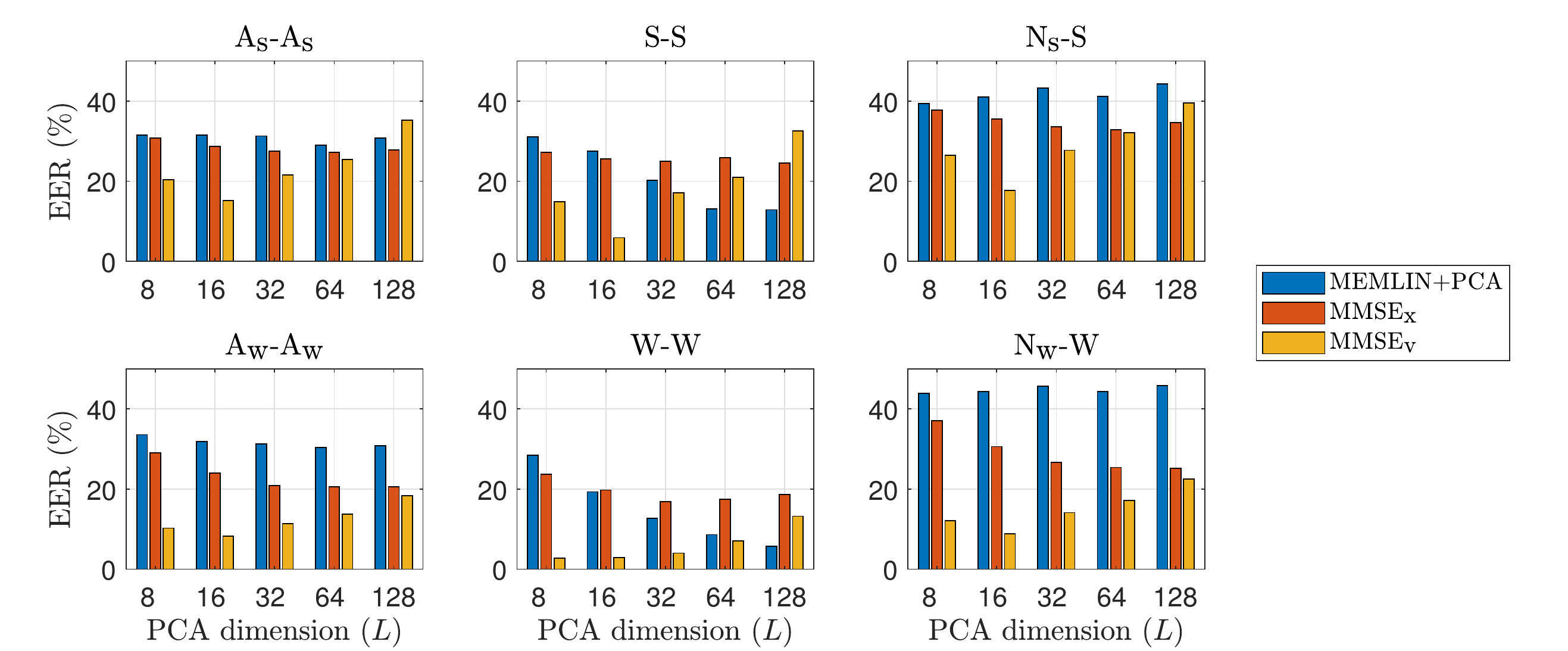}
    \caption{Speaker verification results in terms of EER, in percentages, as a function of the dimensionality, after PCA application, of the embeddings processed by MEMLIN, MMSE$_{\mbox{x}}$ and MMSE$_{\mbox{v}}$. Bar plots are shown for shouted and normal speech (top row), as well as for whispered and normal speech (bottom row).}
    \label{fig:pca_results}
\end{figure*}

\begin{table*}[t]
	\begin{center}
		\caption{Speaker verification results in terms of EER, in percentages, when considering both shouted and normal speech. MEMLIN+PCA, MMSE$_{\mbox{x}}$ and MMSE$_{\mbox{v}}$ process, after PCA application, $L=16$-dimensional embeddings.}
		\label{tab:results_shouted}
		\begin{tabular}{c|cc|c|ccc}
			\toprule
			\textbf{Condition} & \textbf{E-T+MFCC} & \textbf{E-T+WavLM} & \textbf{MEMLIN} & \textbf{MEMLIN+PCA} & \textbf{MMSE$_{\mbox{x}}$} & \textbf{MMSE$_{\mbox{v}}$} \\
			\midrule
			A$_{\mbox{s}}$-A$_{\mbox{s}}$ & 19.96 & 17.11 & 15.62 & 31.50 & 28.72 & \textbf{15.22} \\
			N$_{\mbox{s}}$-N$_{\mbox{s}}$ & 9.73 & \textbf{7.25} & \textbf{7.25} & \textbf{7.25} & \textbf{7.25} & \textbf{7.25} \\
			S-S & 11.58 & 9.94 & 10.44 & 27.46 & 25.53 & \textbf{5.91} \\
			N$_{\mbox{s}}$-S & 25.28 & 21.76 & 20.74 & 41.00 & 35.56 & \textbf{17.74} \\
			\bottomrule
		\end{tabular}
	\end{center}
\end{table*}

\begin{table*}[t]
	\begin{center}
		\caption{Speaker verification results in terms of EER, in percentages, when considering both whispered and normal speech. MEMLIN+PCA, MMSE$_{\mbox{x}}$ and MMSE$_{\mbox{v}}$ process, after PCA application, $L=16$-dimensional embeddings.}
		\label{tab:results_whispered}
		\begin{tabular}{c|cc|c|ccc}
			\toprule
			\textbf{Condition} & \textbf{E-T+MFCC} & \textbf{E-T+WavLM} & \textbf{MEMLIN} & \textbf{MEMLIN+PCA} & \textbf{MMSE$_{\mbox{x}}$} & \textbf{MMSE$_{\mbox{v}}$} \\
			\midrule
			A$_{\mbox{w}}$-A$_{\mbox{w}}$ & 16.54 & 11.24 & \textbf{8.25} & 31.87 & 23.95 & 8.27 \\
			N$_{\mbox{w}}$-N$_{\mbox{w}}$ & 1.21 & \textbf{0.62} & \textbf{0.62} & \textbf{0.62} & \textbf{0.62} & \textbf{0.62} \\
			W-W & 4.38 & 5.26 & 4.00 & 19.31 & 19.77 & \textbf{2.87} \\
			N$_{\mbox{w}}$-W & 12.81 & 9.81 & 11.47 & 44.38 & 30.59 & \textbf{8.86} \\
			\bottomrule
		\end{tabular}
	\end{center}
\end{table*}

\subsection{System Implementation Details}
\label{ssec:details}

The used ECAPA-TDNN back-end was trained, employing the additive angular margin (AAM) loss \cite{AAM19}, on an augmented version of the VoxCeleb2 \cite{VoxCeleb2} dataset to extract $D=256$-dimensional speaker embeddings. Considering an AAM loss margin of 0.2, first, WavLM ---which was pre-trained on 94k hours of unlabeled speech data--- was fixed and the ECAPA-TDNN parameters were trained for a total of 20 epochs. Second, WavLM and the ECAPA-TDNN back-end were jointly fine tuned for 5 epochs. Finally, by following the large margin fine-tuning strategy reported in \cite{Jenthe21}, WavLM and the ECAPA-TDNN back-end were jointly trained for 2 more epochs by considering an AAM loss margin of 0.4. Notice that, for the sake of reproducibility, the model corresponding to this speaker verification system is publicly available\footnote{\url{https://github.com/microsoft/unilm/tree/master/wavlm}}. The reader is referred to \cite{WavLM} for further information on this speaker verification system.

\section{Experimental Results}
\label{sec:results}

%\begin{figure*}
%    \centering
%    \includegraphics[width=0.49\linewidth]{./Figures/Shouted} \hfill \includegraphics[width=0.49\linewidth]{./Figures/Whispered}
%    \caption{Speaker verification results in terms of EER, in percentages, as a function of the dimensionality, after PCA application, of the embeddings processed by MEMLIN, MMSE$_{\mbox{x}}$ and MMSE$_{\mbox{v}}$. Curves are plotted for shouted and normal speech (left), as well as for whispered and normal speech (right).}
%    \label{fig:pca_results}
%\end{figure*}

In this section, EER is chosen as the speaker verification performance metric. Besides, as in previous work \cite{Prieto20,Prieto22}, all the embedding compensation techniques evaluated make use of $K=8$-component GMMs.

\subsection{WavLM Performance}

Tables \ref{tab:results_shouted} and \ref{tab:results_whispered} show speaker verification results in terms of EER under the shouted-normal and whispered-normal scenarios, respectively. The left part of these tables compare, when no embedding compensation is considered, the use of WavLM speech representations (as in Section \ref{sec:overview}), E-T+WavLM, with the use of traditional speech features, E-T+MFCC (note that E-T stands for ECAPA-TDNN). Specifically, the speaker verification system E-T+MFCC, which is publicly available\footnote{\url{https://huggingface.co/speechbrain/spkrec-ecapa-voxceleb}}, employs 80-dimensional Mel-frequency cepstral coefficients \cite{MFCC}. In line with \cite{WavLM}, we can see from these tables that E-T+WavLM generally outperforms E-T+MFCC. That being said, we can also observe that there is still a large room for improvement in the presence of vocal effort mismatch (all conditions except N$_{\mbox{s}}$-N$_{\mbox{s}}$ and N$_{\mbox{w}}$-N$_{\mbox{w}}$) that will be addressed by embedding compensation in the next subsections. Bear in mind that all the embedding compensation experiments in this section are carried out by employing E-T+WavLM as the baseline system.

\subsection{Effect of PCA Dimension}

Figure \ref{fig:pca_results} plots the EER performance of the estimation methodology proposed in Section \ref{sec:mmse}, MMSE$_{\mbox{v}}$, as a function of the PCA dimension $L$. For comparison, these bar plots also show results from MEMLIN (applied in the PCA domain) as well as from an MMSE estimator equivalent to that of Section \ref{sec:mmse} that directly estimates the normal embedding $\tilde{\mathbf{x}}$ from $\mathbb{E}[\mathbf{x}|\mathbf{y}]$, MMSE$_{\mbox{x}}$. From this figure, we can see that MEMLIN's performance tends to drop when decreasing $L$ as a result of the information loss caused by PCA compression, which can be particularly harmful when the estimation relies on a small set of pre-computed and fixed partial estimates.

On the other hand, MMSE$_{\mbox{v}}$ involves the computation of $2L\times 2L$ covariance matrices, $\boldsymbol\Sigma_z^{\{k\}}$, under a data scarcity scenario. Given our small sample size, reducing $L$ helps to achieve better-conditioned covariance matrices to be used in Eqs. (\ref{eq:comb_weights}) and (\ref{eq:partial}). This, together with the fact that MMSE$_{\mbox{v}}$ exploits the observed non-neutrally-phonated embedding $\tilde{\mathbf{y}}$ for partial estimate calculation, can explain why EER decreases up to $L=16$ for MMSE$_{\mbox{v}}$ (see Figure \ref{fig:pca_results}). Keeping decreasing $L$ beyond this point harms speaker verification performance due to the information loss entailed by PCA compression.

In relation to MMSE$_{\mbox{x}}$, an internal analysis revealed that estimating the normal embedding $\tilde{\mathbf{x}}$ from $\mathbb{E}[\mathbf{x}|\mathbf{y}]$ yields target and non-target score probability masses that are poorly separated as a result of compensated embeddings $\hat{\tilde{\mathbf{x}}}$ where the specific-speaker information is significantly distorted. Interestingly, we also observed that the vocal effort transfer vector $\tilde{\mathbf{v}}$ has a weak speaker-dependence. Therefore, estimating $\tilde{\mathbf{x}}$ as $\tilde{\mathbf{y}}-\hat{\tilde{\mathbf{v}}}$ according to MMSE$_{\mbox{v}}$ better preserves the specific-speaker information contained in $\tilde{\mathbf{y}}$, which, in turn, leads to better-separated target and non-target score probability masses.

\subsection{Embedding Compensation Performance Summary}

The right part of Tables \ref{tab:results_shouted} and \ref{tab:results_whispered} compare standard MEMLIN (i.e., without PCA) with MMSE$_{\mbox{v}}$, MMSE$_{\mbox{x}}$ and MEMLIN applied in the PCA domain (MEMLIN+PCA). Note that, in these tables, the three latter techniques process, after PCA application, $L=16$-dimensional embeddings. Under the shouted-normal scenario (Table \ref{tab:results_shouted}), MMSE$_{\mbox{v}}$ outperforms MEMLIN in the presence of vocal effort mismatch (i.e., in A$_{\mbox{s}}$-A$_{\mbox{s}}$, S-S and N$_{\mbox{s}}$-S). Furthermore, while MEMLIN is on par with MMSE$_{\mbox{v}}$ in A$_{\mbox{w}}$-A$_{\mbox{w}}$ under the whispered-normal scenario (Table \ref{tab:results_whispered}), MMSE$_{\mbox{v}}$ achieves in N$_{\mbox{w}}$-W a 22.7\% EER relative improvement with respect to MEMLIN which actually worsens the baseline system E-T+WavLM (as in the S-S condition).

\section{Concluding Remarks}
\label{sec:conclusions}

In this work, we have shown that embedding compensation can significantly mitigate the speaker verification performance drop caused by vocal effort mismatch when a state-of-the-art speaker verification system integrating a cutting-edge self-supervised pre-trained model for speech representation is used. With the aim of improving a reference embedding compensation method ---i.e., MEMLIN---, we have proposed an MMSE estimator of the vocal effort transfer vector that, unlike MEMLIN, exploits the non-neutrally-phonated embeddings observed at test time for partial estimate calculation and performs in a PCA domain to cope with non-neutrally-phonated speech data scarcity. Compared with MEMLIN, the proposed MMSE estimator has shown superior and competitive EER performance when processing shouted and whispered speech, respectively.

% Below is an example of how to insert images. Delete the ``\vspace'' line,
% uncomment the preceding line ``\centerline...'' and replace ``imageX.ps''
% with a suitable PostScript file name.
% -------------------------------------------------------------------------
%\begin{figure}[htb]

%\begin{minipage}[b]{1.0\linewidth}
%  \centering
%  \centerline{\includegraphics[width=8.5cm]{image1}}
%  \vspace{2.0cm}
%  \centerline{(a) Result 1}\medskip
%\end{minipage}
%
%\begin{minipage}[b]{.48\linewidth}
%  \centering
%  \centerline{\includegraphics[width=4.0cm]{image3}}
%  \vspace{1.5cm}
%  \centerline{(b) Results 3}\medskip
%\end{minipage}
%\hfill
%\begin{minipage}[b]{0.48\linewidth}
%  \centering
%  \centerline{\includegraphics[width=4.0cm]{image4}}
%  \vspace{1.5cm}
%  \centerline{(c) Result 4}\medskip
%\end{minipage}
%
%\caption{Example of placing a figure with experimental results.}
%\label{fig:res}
%
%\end{figure}

\bibliographystyle{IEEEbib}
\bibliography{strings,refs}

\end{document}